\documentclass[reprint,superscriptaddress,amsmath,amssymb,aps,prl]{revtex4-2}

\usepackage{braket} 
\usepackage{amssymb}
\usepackage{amsmath} 
\usepackage{amsfonts}

\usepackage{bm}
\usepackage{booktabs}  
\usepackage{braket}

\usepackage{caption}
\usepackage{comment}

\usepackage{dcolumn}

\usepackage{epsfig}
\usepackage{epstopdf}
\usepackage{enumitem} 

\usepackage{float}

\usepackage{graphicx}

\usepackage{hyperref}

\usepackage{ulem}

\usepackage{xcolor}

\usepackage[justification=raggedright]{caption} 

\UseRawInputEncoding

\begin{document}
\title{Demonstration of Discrete-Time Quantum Walks and Observation of Topological Edge States in a Superconducting Qutrit Chain}

\author{Kun Zhou\textsuperscript{\textasteriskcentered}}
\affiliation{National Laboratory of Solid State Microstructures, School of Physics, Nanjing University, Nanjing 210093, China}
\affiliation{School of Physics, Hangzhou Normal University, Hangzhou, 	Zhejiang 311121, China}
\affiliation{Hefei National Laboratory, Hefei 230088, China}
\affiliation{Shishan Laboratory, Suzhou Campus of Nanjing University, Suzhou 215000, China}

\author{Jian-Wen Xu\textsuperscript{\textasteriskcentered}}
\affiliation{National Laboratory of Solid State Microstructures, School of Physics, Nanjing University, Nanjing 210093, China}

\author{Qi-Ping Su\textsuperscript{\textasteriskcentered,\textdagger}}
\affiliation{School of Physics, Hangzhou Normal University, Hangzhou, 	Zhejiang 311121, China}

\author{Yu Zhang}
\affiliation{National Laboratory of Solid State Microstructures, School of Physics, Nanjing University, Nanjing 210093, China}
\affiliation{Hefei National Laboratory, Hefei 230088, China}
\affiliation{Shishan Laboratory, Suzhou Campus of Nanjing University, Suzhou 215000, China}

\author{Xiang-Min Yu}
\affiliation{National Laboratory of Solid State Microstructures, School of Physics, Nanjing University, Nanjing 210093, China}
\affiliation{Hefei National Laboratory, Hefei 230088, China}
\affiliation{Shishan Laboratory, Suzhou Campus of Nanjing University, Suzhou 215000, China}

\author{Zhuang Ma}
\affiliation{National Laboratory of Solid State Microstructures, School of Physics, Nanjing University, Nanjing 210093, China}

\author{Han-Yu Zhang}
\affiliation{National Laboratory of Solid State Microstructures, School of Physics, Nanjing University, Nanjing 210093, China}

\author{Hong-Yi Shi}
\affiliation{National Laboratory of Solid State Microstructures, School of Physics, Nanjing University, Nanjing 210093, China}

\author{Wen Zheng}
\affiliation{National Laboratory of Solid State Microstructures, School of Physics, Nanjing University, Nanjing 210093, China}
\affiliation{Shishan Laboratory, Suzhou Campus of Nanjing University, Suzhou 215000, China}

\author{Shuyi Pan}
\affiliation{National Laboratory of Solid State Microstructures, School of Physics, Nanjing University, Nanjing 210093, China}

\author{Yihao Kang}
\affiliation{School of Physics, Hangzhou Normal University, Hangzhou, 	Zhejiang 311121, China}

\author{Zhiguo Huang}
\affiliation{China Mobile (Suzhou) Software Technology Company Limited, Suzhou, 215163, China}

\author{Chui-Ping Yang\textsuperscript{\textdaggerdbl}}
\affiliation{School of Physics, Hangzhou Normal University, Hangzhou, 	Zhejiang 311121, China}

\author{Shao-Xiong Li\textsuperscript{\textsection}}
\affiliation{National Laboratory of Solid State Microstructures, School of Physics, Nanjing University, Nanjing 210093, China}
\affiliation{Hefei National Laboratory, Hefei 230088, China}
\affiliation{Shishan Laboratory, Suzhou Campus of Nanjing University, Suzhou 215000, China}
\affiliation{Jiangsu Key Laboratory of Quantum Information Science and Technology, Nanjing University, Suzhou 215163, China}

\author{Yang Yu\textsuperscript{\textparagraph}}
\affiliation{National Laboratory of Solid State Microstructures, School of Physics, Nanjing University, Nanjing 210093, China}
\affiliation{Hefei National Laboratory, Hefei 230088, China}
\affiliation{Shishan Laboratory, Suzhou Campus of Nanjing University, Suzhou 215000, China}
\affiliation{Jiangsu Key Laboratory of Quantum Information Science and Technology, Nanjing University, Suzhou 215163, China}
\affiliation{Synergetic Innovation Center of Quantum Information and Quantum Physics, University of Science and Technology of China, Hefei, Anhui 230026, China}

\date{\today}
\begin{abstract}
Quantum walk serves as a versatile tool for universal quantum computing and algorithmic research. 
However, the implementation of discrete-time quantum walks (DTQWs) with superconducting circuits is still constrained by some limitations such as operation precision, circuit depth and connectivity. With improved hardware efficiency by using superconducting qutrits (three-level systems), we experimentally demonstrate a scalable DTQW in a superconducting circuit, observing the ballistic spreading of quantum walk in a qutrit chain.
The usage of qutrits in our implementation allows hardware efficiently encoding of the walker position and the coin degree of freedom. 
By exploiting the flexibility and intrinsic symmetries of qutrit-based DTQWs, we successfully prepare two topological phases in the chain. 
For the first time, particle-hole-symmetry-protected edge states, bounded at the interface between these two topological phases, are observed in the superconducting platform. Measured parameter dependencies further validate the properties of edge states. 
The scalability and gate-control compatibility of the demonstrated DTQWs enable a versatile tool for superconducting quantum computing and quantum simulation.
\end{abstract}

\maketitle

Quantum walks (QWs) constitute a foundational model in the study of quantum information processing. 
As a foundational algorithmic tool, QW has been extensively employed in the design of diverse quantum algorithms \cite{q_alg_1,q_alg_3,q_alg_4}.  Particularly, some features of QWs are very instrumental. 
For instance, the coherent dynamics of QWs can significantly accelerate search algorithms \cite{Graph_2,QW_search_1,QW_search_2}. Notably, QWs have been proven to traverse specific structures, such as random hierarchical graphs, exponentially faster than classical algorithms \cite{QW_EXPSUP}. Furthermore, the distinctive probability distribution of QWs can be used to efficiently implement boson sampling protocols \cite{BS_2019}.
In addition, QWs by themselves can carry out universal quantum computation \cite{u_c_1,u_c_2,u_c_3} and quantum simulation \cite{q_sim_1,q_sim_2,q_sim_3}. 
Existing demonstrations of QWs in superconducting systems encompass the two predominant models: continuous-time and discrete-time approaches. 
The former \cite{qw_in_sc_2_PJW1D,qw_in_sc_3_PJW2D} generally works in an analogy way and thus does not fully align with the gate-based quantum computing paradigm. 
The unitary evolution of the latter comprises a coin toss and a subsequent coin-state-dependent shift of the walker, which possesses higher controllability and can be encoded into gate sequences in quantum circuits. 
Nevertheless, the circuit depth and connectivity required for qubit-based DTQWs \cite{DTQW_circuit} remains challenging. 
The implementation reported in Ref. \cite{qw_in_sc_1_phase_space,qw_in_sc_2_phase_space} only requires a qubit acting as the coin and a cavity encoding the walker's position, while bringing in some difficulties in scalability (the ability to scale up efficiently in walk step count and walk dimensionality) and flexibility. 
Up to now, the implemented DTQWs in superconducting systems appear to not be able to achieve accessibility and scalability within the quantum circuit framework, limiting the practical applications of DTQWs. 

Going beyond two-level systems (qubits), multilevel quantum systems (qudits) can bring in many benefits such as  the optimized multi-qubit encoding and gate decomposition in circuit compilation \cite{qudit_encode0,qudit_encode1,qudit_encode2,qudit_compile1,qudit_compile2,qudit_compile3,qudit_compile4}, the more efficient implementation of quantum algorithms \cite{Qudit,qudit_search,qudit_shor,qudit_yhf}, and inherent advantages in quantum error correction \cite{qudit_errorcorrect1,qudit_errorcorrect2}. 
These benefits have motivated the extensive study of qudit-based quantum technologies in photonics \cite{qudit_photonics1,qudit_photonics2}, neutral atoms and ions \cite{qudit_ions,qudit_Rydberg}, and superconducting  platforms \cite{qudit_yhf,qudit_errormitigation}. In particular, gate operation \cite{qudit_yhf} and logical-qubit construction \cite{qudit_errorcorrect2} based on qudits recently have made significant progress in superconducting circuit systems. 
In this Letter, we experimentally demonstrate that qutrits (three-level quantum systems) can be used to facilitate the  implementation of discrete-time quantum walks (DTQWs) with improved hardware efficiency, operation flexibility and scalability in a superconducting circuit system. 
The uniqueness of our qutrit-based DTQW protocol lies in its inherent integration of the coin at each walker position, eliminating the need for a dedicated coin qubit. This design significantly simplifies the circuit implementation and improves the system controllability, despite the challenges intrinsic to superconducting platforms.

\begin{figure*}[th]
	\includegraphics[scale=0.32]{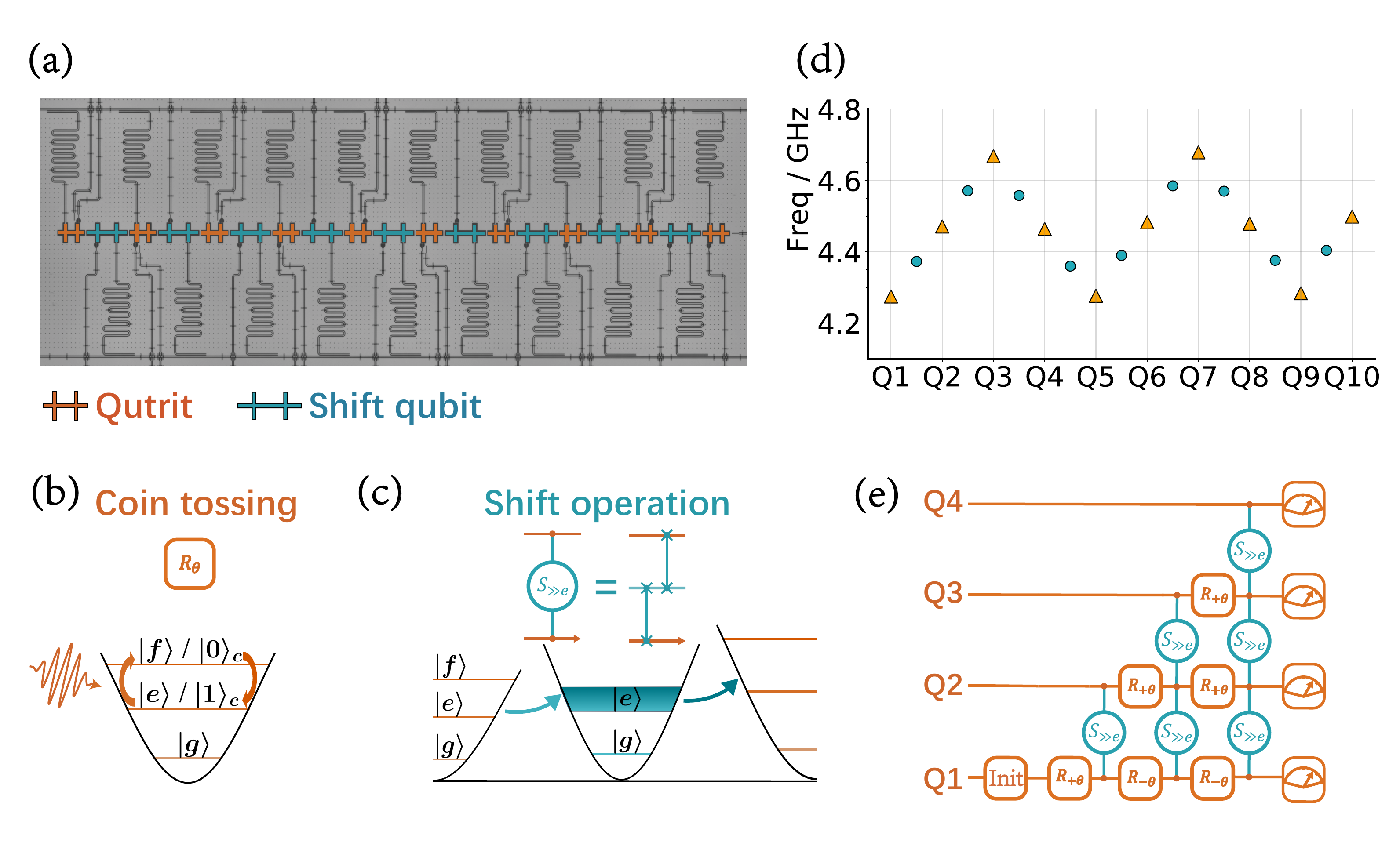}
	\caption{\label{fig1} \textbf{DTQW in a 1D chain of a superconducting chip.} \textbf{a,} Optical micrograph of the 19-transmon chain. Each qutrit is equipped with an independent XY control line and flux bias line, except for the leftmost qutrit, which only has a flux bias line. Each shift qubit (SQ) is equipped with a flux bias line. Both the qutrits and the SQs are coupled to a transmission line through their own $\lambda/4$ readout resonator for readout measurement. 
	\textbf{b,} The coin operation $R_{\theta}$. The tossing is executed by an SU(2) gate within the subspace $\left\{\ket{e}, \ket{f}\right\}$ of the qutrit. The subscript $\theta$ denotes the rotation angle about an axis in the equatorial plane.
	\textbf{c,} The shift operation $S_{\gg e}$. This operation resembles the operation of a classical shift register, moving the $\ket{e}$ state to the right by one position. It is realized through two sequential swap operations, which are executed by frequency tuning of an SQ. The extended width of the $\ket{e}$ state of the SQ indicates the frequency tuning. 
	\textbf{d,} The zig-zag spatial distribution of the frequency of the transmons. $Q_i$ labels $i$th qutrit (orange triangles), and blue circles are SQs. 
	\textbf{e,} An effective quantum logical circuit for a 3-step DTQW. In the $n$-th step of the walk, coin tosses performed synchronously on the first $n$ qutrits constitute the coin operation described by Eq.~(\ref{coin}). Similarly, synchronous SWAP pairs constitute the shift operation described by Eq.~(\ref{walker}). }
\end{figure*}
 
We demonstrate the flexible controllability of the qutrit-based DTQWs by experimentally generating  two distinct topological phases in a superconducting 10-qutrit chain. 
The topological edge states are observed at the interface between the two topological phases, verifying the theoretical prediction \cite{prb2012,pra2023}. The existence of the edge states in our implemented DTQWs is further confirmed through measured parameter dependencies. 
Notably, the protection mechanism of these boundary states is attributed to particle-hole symmetry (PHS), which differs fundamentally from those reported previously \cite{qw_in_sc_1_phase_space,splitstep,a4,topo2D_xue,pra2010}. 
Potentially, our qutrit-based protocol propels QWs into practical applications. 

\emph{Protocol}.---DTQWs constitute a quantum system whose Hilbert space can be decomposed into a coin space $H_c$ and a walker position space $H_s$. For one dimensional (1D) DTQWs, an arbitrary state in the space $H_c\otimes H_s$ is described by $\ket{\phi}=\sum_x (a_x\ket{0}+b_x\ket{1})\otimes \ket{x}$, where $x\in \textbf{Z}$ is the position of the walker and $\ket{0}$, $\ket{1}$ are two basis states of the coin. The discrete time-step ($t$) evolution $U$ of the system consists of two unitary operators: the shift operator $S$ and the coin operator $R$. A single step of the DTQW can be expressed as 
	\begin{equation}
		\ket{\phi(t+1)}=U\ket{\phi(t)}=SR\ket{\phi(t)}.
	\end{equation}
The operator $S$ conditionally shifts the walker from one position to its neighboring position:
    \begin{equation}
    	S=\ket{0}\bra{0}\otimes I+
    	\sum_{x}\ket{1}\bra{1}\otimes\ket{x+1}\bra{x}. \label{walker}
    \end{equation}
Note that the above walk is a compact unidirectional walk. Its dynamics is mathematically equivalent to that of the traditional bidirectional walk, where the first term of $S$ is replaced as $\sum_{x}\ket{0}\bra{0}\otimes\ket{x-1}\bra{x}$. Consequently, these two types of walks can be mapped between each other \cite{UD_1,UD_2} (See Supplementary Information \cite{SupplementalMaterial} for detailed comparison and converting method).
The operator $R$ is a unitary rotation that tosses the coin:
	\begin{equation}
		R=\sum_x R(x)\otimes \ket{x}\bra{x}.
		\label{coin}
	\end{equation}
For a rotation in two-level systems, without loss of generality, $R(x)=$ exp$[i\theta(x)\sigma_x/2]$, where $\sigma_x$ is the $x$ component of the Pauli operator and $\theta$ is the position-dependent rotation angle.

\begin{figure*}[th]
	\includegraphics[scale=0.32]{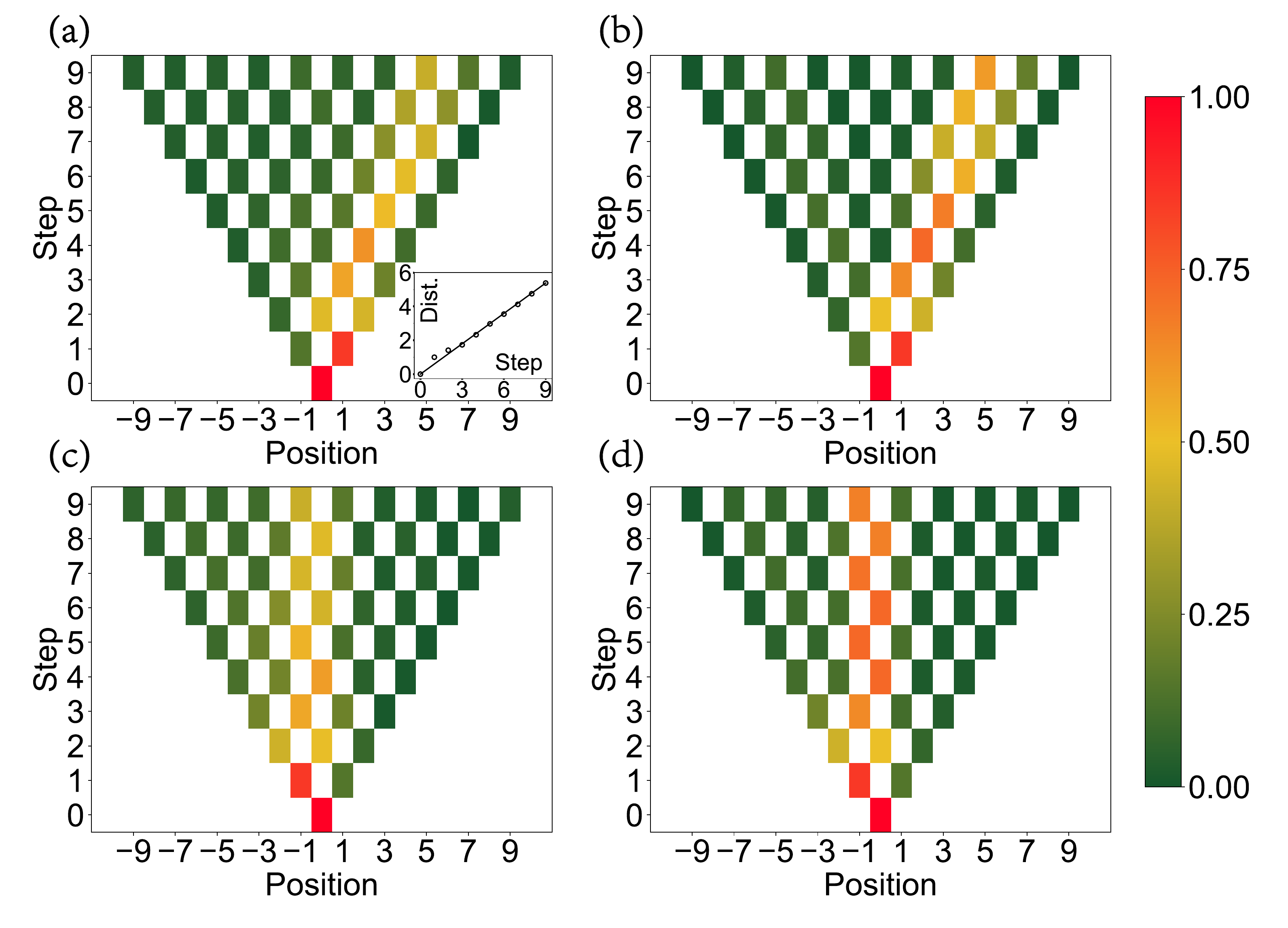}
	\caption{\label{fig2} \textbf{DTQW results for edge state and non-edge state.} \textbf{a,} The measured excitation distribution of DTQWs with $\theta_\pm=\pm\pi/4$ and the initial state $\ket{\phi_{co}^0}$. The inset is the obtained walk step dependence of the diffusion distance $D(t)=\sqrt{\sum_x x^2 p(x,t)}$, where $p(x,t)$ is the population  (equivalent to probability) at position $x$. \textbf{b,} The calculated results for $\theta_\pm=\pm\pi/4$ and the initial state $\ket{\phi_{co}^0}$ without considering gate errors. \textbf{c,} The measured excitation distribution of DTQWs with $\theta_\pm=\pm\pi/4$ and the initial state $\ket{\phi_{ce}^0}$. \textbf{d,} The calculated results for $\theta_\pm=\pm\pi/4$ and the initial state $\ket{\phi_{ce}^0}$ without considering gate errors.
	}
\end{figure*}

In implementing DTQWs with a superconducting circuit system, our primary concern is making the system's architecture scalable and operations efficient. The widely adopted transmon \cite{transmon} is used as the building block. Our 1D DTQWs system is formed by an 1D array of 19 transmons, where 10 out of 19 transmons function as qutrits, and the other 9 transmons function as shift qubits (SQs). The qutrits and SQs alternate as shown in Fig.~\ref{fig1}a. There is a capacitive coupling between the neighboring transmons. 
To facilitate the execution of operator $S$ in the entire chain and minimize crosstalks, a triangular frequency distribution is arranged for the transmons as shown in Fig.~\ref{fig1}d. While this type of direct-coupling architecture has limitations in aspects such as gate fidelity and scalability, it is sufficient for demonstrating our protocol.
The coupler-based architecture \cite{cp_1,cp_2,cp_3,cp_4}, although not being employed in this work, is also feasible for our DTQW protocol, which would require only minor modifications to the implementation of operator $S$. This coupler-based architecture is advantageous in terms of stray interaction suppression, gate fidelity, and task flexibility. In addition, freed from the trade-off between gate speed and stability of higher excited states, it exhibits the potential to support quantum walks at a larger scale (see Supplementary Information \cite{SupplementalMaterial} for the discussion on compatibility and scalability of our DTQW protocol).

In our DTQWs implementation, the usage of qutrit, as shown in Fig.~\ref{fig1}b, enables the coin state and a specific walker position simultaneously to be encoded on a single transmon qutrit: 
its ground state $\ket{g}$ encodes an unoccupied walker position, while the two distinct excited states $\ket{e},\ket{f}$ encode both walker present position and the coin degree of freedom. 
Compared with the qubit-based schemes (e.g., Ref. \cite{DTQW_2qbit}), which require two superconducting qubits at each position to encode the joint Hilbert space of the walker's occupation states (present or absent) and the coin states, our implementation requires fewer tranmons and thus improves hardware efficiency. Moreover, the position-dependent $R$ operation is performed by a SU(2) gate in the subspace $\left\{\ket{e}, \ket{f}\right\}$, obviating the need for two-transmon interactions as in Refs. \cite{DTQW_2qbit}. 
Regarding the implementation of the shift operator, 
the tunable SQ completes shift operation as shown in Fig.~\ref{fig1}c. When the SQ is tuned to be on resonance with the $\ket{g}\leftrightarrow\ket{e}$ transition of the neighboring qutrit, SWAP gates can shift the excitation of $\ket{e}$ into and out of SQs with no shift for the $\ket{f}$ state. Two sequential SWAP gates complete one step forward of excitation $\ket{e}$ at any position, i.e., realizing the S operation. 
Evidently, this approach both simplifies operator compilation and reduces physical connectivity requirements compared with the qubit-based protocol in Ref. \cite{DTQW_circuit}. 

As an example, a quantum circuit for a 3-step DTQW is presented in Fig.~\ref{fig1}e. The circuit begins by initializing the walker and coin states through, respectively, a $\pi_{\text{ge}}$ pulse and an SU(2) gate in the $\{\ket{e}, \ket{f}\}$ subspace. 
For the $n$-th step evolution, the coin operation comprises $n$ parallel SU(2) gates on the first $n$ qutrits, while the shift operation consists of $n$ corresponding shift blocks. 
The readout measurements only require extracting the excitation number of the qutrits, which is a relatively easy task. 
Note that this protocol is scalable -- the walk scale can be increased by simply extending the qutrit chain length, and can likewise be generalized to two-dimensional (2D) DTQW implementations without changing the encoding scheme and the implementation of the operators \cite{2Dwalk} (see Supplementary Information \cite{SupplementalMaterial} for the discussion on compatibility and scalability of our DTQW protocol).

\emph{Results.}---Since DTQWs are an excellent platform for studying topological phases \cite{topo2D_xue,topo_xue,q_sim_2,cold_atoms,qw_in_sc_1_phase_space}, we generate topological edge states to test the effectiveness of our hardware design and operation method. 
As described earlier, DTQWs are realized by repeatedly applying $U$ on the chain. The degrees of freedom of coin in $U$ can be viewed as the spin degree of freedom, and the degree of freedom of walker can be seen as the orbital degree of freedom. DTQW thus has a spin-orbit coupling, which is the key to generate topological phases, while CTQW does not have this type of coupling. 
It has been numerically shown that the parameter space of DTQW can have two separated domains, which allow two distinct topological phases \cite{prb2012,pra2023}. The topology arises from the PHS and the pairwise of the edge states is attributed to the sublattice symmetry of DTQWs \cite{prb2012}.
We construct inhomogeneous DTQWs with
\begin{equation}\label{eq1}
	\theta(x) = 
	\begin{cases} 
		\theta_-<0 & \text{for}\ x < 0, \\
		\theta_+ \ge 0 & \text{for}\ x \geq 0. 
	\end{cases}
\end{equation}
In the DTQWs, regions with $x<0$ and $x \ge 0$ correspond to two distinct topological phases. 
At the interface of the two topological phases, there exist two edge states:
\begin{equation}
	\ket{\phi_e}=\frac{1}{N}\sum_x e^{i\omega x}\ket{x}\otimes(a_x\ket{1}+ib_x\ket{0}),
\end{equation}
where $\omega=0$ and $\pi$ for the two edge states, respectively; $a_x=\left[(1-\sin{\theta(x)})/\cos{\theta(x)}\right]^x$; $b_x=\left[(1-\sin{\theta(x)})/\cos{\theta(x)}\right]^{x+1}$; and the normalization constant $N=(1/\sin{\theta_+}-1/\sin{\theta_-})^{1/2}$.

\begin{figure}[h]
	\includegraphics[scale=0.25]{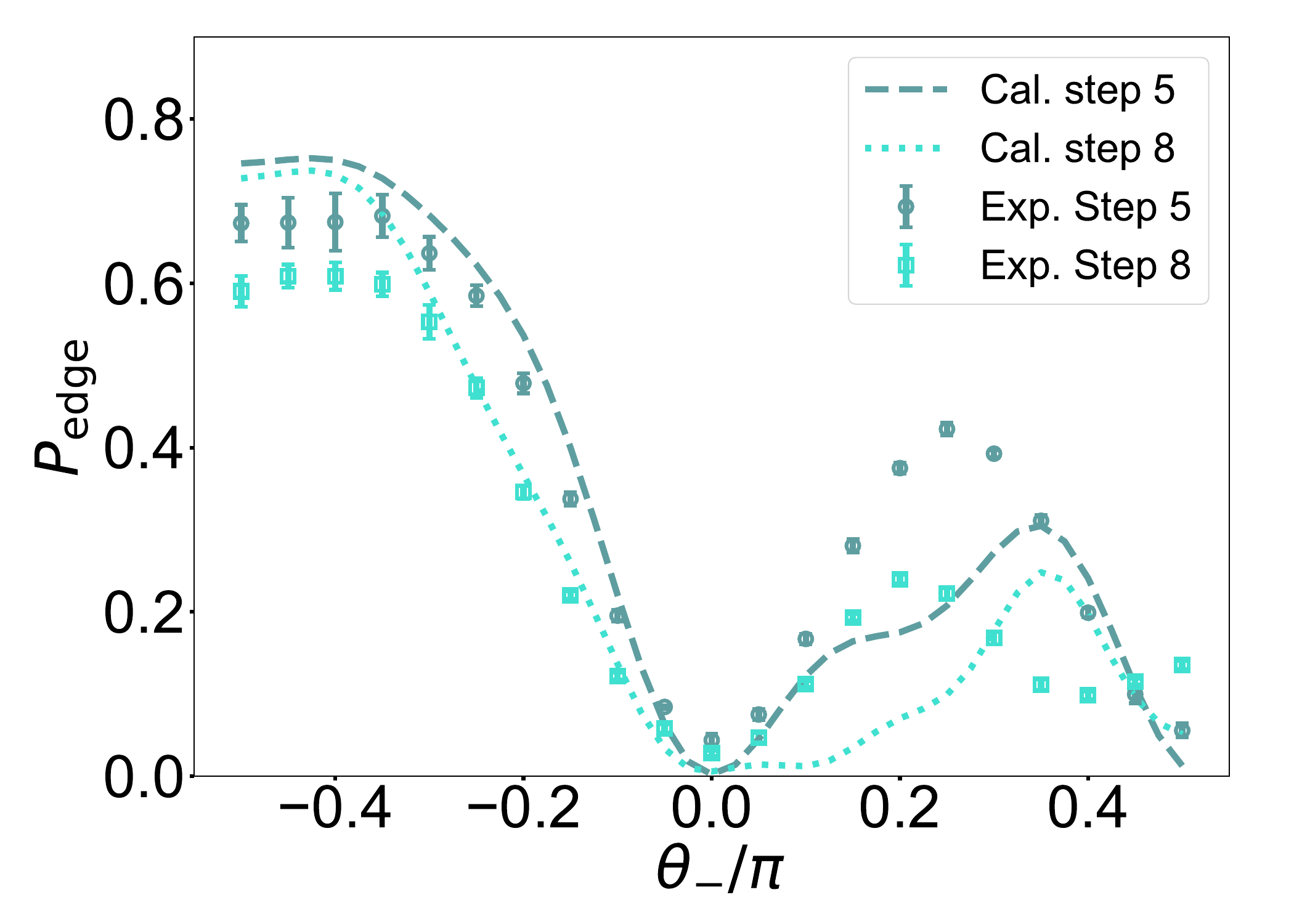}
	\caption{\label{fig3} Populations ($P_{\text{edge}}$) around $x=0$ with the initial state at $\ket{\phi_{ce}^0}$, fixed $\theta_{+}=\pi/4$ and varying $\theta_{-}$, for 5-step (grayish-blue) and 8-step (cyan) DTQWs respectively. Symbols are experimental data (circles for 5-step DTQWs; squares for 8-step DTQWs), and lines are theoretical calculation (dashed line for 5-step DTQWs; dotted line for 8-step DTQWs). Error bars denote the standard deviation of seven replicates.} 
\end{figure}

\begin{figure}[h]
	\includegraphics[scale=0.25]{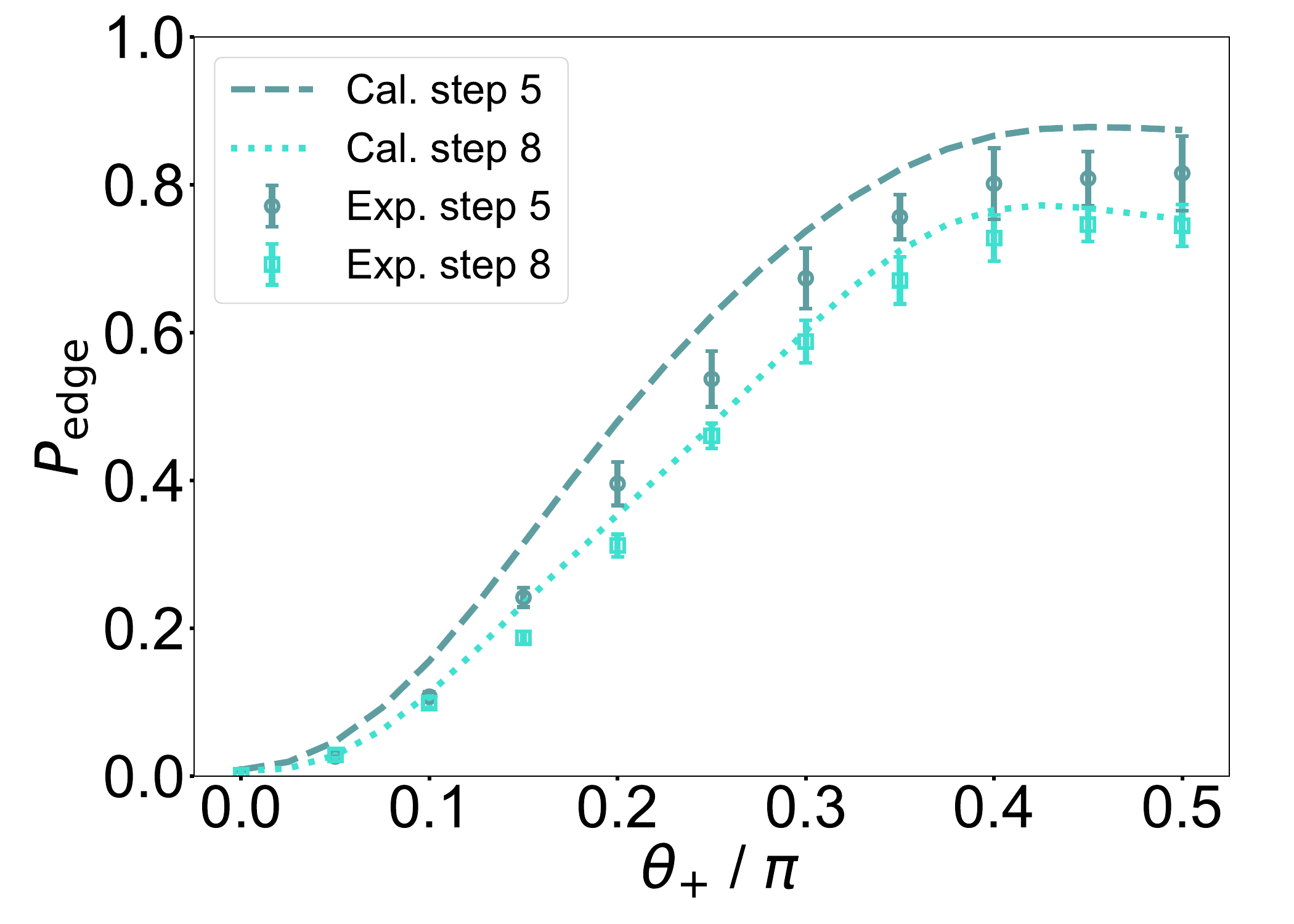}
	\caption{\label{fig4} Populations ($P_{\text{edge}}$) around $x=0$ with the initial state at $\ket{\phi_{ce}^0}$, fixed relationship $\theta_{+}=-\theta_{-}$, and varying $\theta_{+}$ for 5-step (grayish-blue) and 8-step (cyan) DTQWs, respectively. Symbols are experimental data (open circles for 5-step DTQWs; open squares for 8-step DTQWs), and lines are theoretical calculation (dashed line for 5-step DTQWs;  dotted line for 8-step DTQWs). Numerical simulations confirm that the behavior for $\theta_{+} < 0$ is symmetric with respect to $\theta_{+} > 0$, and thus it is not shown. Error bars denote the standard deviation of seven replicates}
\end{figure}

The obtained results of DTQWs are sensitive to the initial coin state. Fig.~\ref{fig2}a shows the measured distributions of DTQWs with $\theta_\pm=\pm\pi/4$ and the initial state:  $\ket{\phi_{co}^0}=\ket{x=0}\otimes[i(\sqrt{2}-1)\ket{1}+\ket{0}]/\sqrt{4-2\sqrt{2}}$. For ease of comparison, the compact one-way walk data are converted to conventional two-way walk data (See Supplementary Information \cite{SupplementalMaterial} for the raw data and converting method). 
Since $\ket{\phi_{co}^0}$ is orthogonal to the edge states $\ket{\phi_e}$, the walker exhibits typical QW behavior in Fig.~\ref{fig2}a: the distribution of DTQWs spreads ballistically from the original position to the right side. 
The spreading distance is proportional to the walk step number $t$ as shown in the inset of Fig.~\ref{fig2}a, in stark contrast to classic random walk's proportion of $\sqrt{t}$. Fig.~\ref{fig2}b shows the  results of theoretical calculation with the same $\theta_\pm$ values and the same initial state. It is obvious that the overall pattern of the experiment results is identical to the calculated distribution pattern. The amplitude difference between these two is mainly due to the infidelity of gate operations. 
We estimate the agreement between experimental results and theoretical calculation by calculating the similarity \cite{q_sim_1,q_sim_3} $Sim(t)=[\sum_x \sqrt{p_{\text{theor}}(x,t) p_{\text{expt}}(x,t)}]^2$, in which $p_{\text{theor}}(x,t)$ and $p_{\text{expt}}(x,t)$ represent the theoretical and experimental population densities at position $x$ for the $t$-th step, respectively. For our 9 step DTQW, $Sim(9) \sim 84.5\%$. 
For the initial state $\ket{\phi_{ce}^0}=\ket{x=0}\otimes[\ket{1}+i(\sqrt{2}-1)\ket{0}]/\sqrt{4-2\sqrt{2}}$ with $\theta_\pm=\pm\pi/4$, which approximately equals the uniform superposition of the two edge states $\sqrt{2}/2( \ket{\phi_e^{\omega=0}}+\ket{\phi_e^{\omega=\pi}} )$ , the experimental and calculated distributions of DTQWs are presented in Fig.~\ref{fig2}c and Fig.~\ref{fig2}d respectively. Again, theoretical calculation and experiment results agree with each other, and $Sim(9) \sim 87.2\%$ is obtained. Since $\ket{\phi_{ce}^0}$ significantly overlaps with the edge states $\ket{\phi_e}$, the obtained distributions of DTQWs concentrate around the interface $x=0$ and do not spread out with the increment of walking steps. 
Note that the distribution bounces back and forth between $x=-1$ and $x=0$. This occurs because at even steps, the two edge states (the main components of the initial state $\ket{\phi_{ce}^0}$) gain zero relative phase and interfere constructively at $x=0$, while at odd steps, they acquire a $\pi$ phase shift and show constructive interference at $x=-1$.
The observed trapping at the interface and the bouncing between two positions strongly suggest that the topological edge states are generated in our gate-controlled DTQWs.

To further confirm the topological edge states in our DTQWs, we examine different $\theta_\pm$ conditions with the same initial state $\ket{\phi_{ce}^0}$. 
We keep $\theta_+$ fixed at $\pi/4$, and change $\theta_-$ from negative values to positive values. Since topological edge states only exist at the interface between two distinct topological phases, it can be clearly seen from Fig.~\ref{fig3} that the probability at the interface for $\theta_-<0$ is larger than that for $\theta_->0$. This observation is consistent with the higher occupation probability of the edge states at the interface. To give a quantitative comparison, we numerically calculate the results by considering the average errors due to gate fidelity and qutrit lifetime (see Supplementary Information \cite{SupplementalMaterial} for specific parameters and error analysis). 
Compared with the calculated results, the experiment data have a large fluctuation in the region of $\theta_{-}>0$ and a relatively small fluctuation in the region of $\theta_{-}<0$. This result can be attributed to the fact that there is a topological protection of the edge states in the region of $\theta_{-}<0$, which makes the system more robust against the experimental errors, while no topological protection exists in the region of $\theta_{-}>0$.

Changing $\theta_\pm$ can tune the edge states' locality. As a result, the overlap of the edge states with the initial state $\ket{\phi_{ce}^0}$ is controllable. We keep $\theta_+=-\theta_->0$, and change the value of $\theta_+$. With $\theta_+$ changed from $0$ to $\pi/2$, the edge states are least localized for $\theta_+=0$, and continuously transit to a completely localized state for $\theta_+=\pi/2$. The overlapping probability $P_0=\sum_\omega\langle\phi_e(\omega)|\phi^0_{ce}\rangle=[2\tan{\theta_+}\ (1-\sin{\theta_+})/\cos{\theta_+}]^2$ protected around $x=0$ can reveal the locality of the edge states. In Fig.~\ref{fig4}, the observed $\theta_+$ dependence of $P_0$ agrees with theoretical prediction, where the errors from the gate operations and the finite lifetime of the qutrits have been included in the theoretical calculation.

\emph{Conclusion.}---Using a superconducting qutrit chain, we have successfully realized the qutrit-based DTQW protocol and observed the characteristic QW ballistic spreading. 
The qutrit encoding allows position-dependent coin operations. Through this ability, we have generated the edge states bounded at the interface between two distinct topological phases in DTQWs. 
The observed $\theta_\pm$ dependency and locality of the edge states both confirm theoretical predictions. 

Our work resolves a longstanding experimental challenge in physical realization of scalable DTQWs within superconducting circuit architectures, enhancing superconducting processors' capabilities for quantum algorithms and simulations. 
The DTQW implementation scales via qutrit chain extension and readily extends to 2D superconducting qubit networks. 
Our experiment presents the first demonstration of PHS-protected edge states via DTQWs, expanding the capacity of superconducting platform to engineer topologically protected quantum states. 
This work also opens an avenue for future study of DTQWs and exploration of quantum effects in engineered high-dimensional systems.

\emph{Acknowledgments.}---This work was partially supported by the Innovation Program for Quantum Science and Technology (Grant No. 2021ZD0301702), the National Natural Science Foundation of China (NSFC) ((Grant Nos. 12074179 and U21A20436), the National key Research and Development program of China (Grant No. 2024YFA1408900), the Natural Science Foundation of Jiangsu Province, China (Grant Nos. BE2021015-1 and BK20232002), the Natural Science Foundation of Shandong Province (Grant No. ZR2023LZH002) and Nanjing University-China Mobile Communications Group Co., Ltd Joint Institute. 

\quad \\
\textsuperscript{\textasteriskcentered}\quad These authors contributed equally to this work. \\
\textsuperscript{\textdagger}\quad Contact author: sqp@hznu.edu.cn. \\
\textsuperscript{\textdaggerdbl}\quad Contact author:yangcp@hznu.edu.cn. \\
\textsuperscript{\textsection}\quad Contact author:shaoxiong.li@nju.edu.cn. \\
\textsuperscript{\textparagraph}\quad Contact author:yuyang@nju.edu.cn.

\bibliography{reference1228}

@article{Qudit,
  title = {Colloquium: Qudits for decomposing multiqubit gates and realizing quantum algorithms},
  author = {Kiktenko, Evgeniy O. and Nikolaeva, Anastasiia S. and Fedorov, Aleksey K.},
  journal = {Rev. Mod. Phys.},
  volume = {97},
  issue = {2},
  pages = {021003},
  numpages = {26},
  year = {2025},
  month = {Jun},
  publisher = {American Physical Society},
  doi = {10.1103/RevModPhys.97.021003},
  url = {https://link.aps.org/doi/10.1103/RevModPhys.97.021003}
}

@article{qudit_search,
  title = {Time-efficient implementation of quantum search with qudits},
  author = {Ivanov, S. S. and Tonchev, H. S. and Vitanov, N. V.},
  journal = {Phys. Rev. A},
  volume = {85},
  issue = {6},
  pages = {062321},
  numpages = {5},
  year = {2012},
  month = {Jun},
  publisher = {American Physical Society},
  doi = {10.1103/PhysRevA.85.062321},
  url = {https://link.aps.org/doi/10.1103/PhysRevA.85.062321}
}

@article{qudit_shor,
  title = {Factoring with qutrits: Shor's algorithm on ternary and metaplectic quantum architectures},
  author = {Bocharov, Alex and Roetteler, Martin and Svore, Krysta M.},
  journal = {Phys. Rev. A},
  volume = {96},
  issue = {1},
  pages = {012306},
  numpages = {17},
  year = {2017},
  month = {Jul},
  publisher = {American Physical Society},
  doi = {10.1103/PhysRevA.96.012306},
  url = {https://link.aps.org/doi/10.1103/PhysRevA.96.012306}
}

@article{qudit_yhf,
  title = {Performing $\mathrm{SU}(d)$ Operations and Rudimentary Algorithms in a Superconducting Transmon Qudit for $d=3$ and $d=4$},
  author = {Liu, Pei and Wang, Ruixia and Zhang, Jing-Ning and Zhang, Yingshan and Cai, Xiaoxia and Xu, Huikai and Li, Zhiyuan and Han, Jiaxiu and Li, Xuegang and Xue, Guangming and Liu, Weiyang and You, Li and Jin, Yirong and Yu, Haifeng},
  journal = {Phys. Rev. X},
  volume = {13},
  issue = {2},
  pages = {021028},
  numpages = {21},
  year = {2023},
  month = {May},
  publisher = {American Physical Society},
  doi = {10.1103/PhysRevX.13.021028},
  url = {https://link.aps.org/doi/10.1103/PhysRevX.13.021028}
}

@article{qudit_encode0,
   title={Emulating two qubits with a four-level transmon qudit for variational quantum algorithms},
   volume={9},
   ISSN={2058-9565},
   url={http://dx.doi.org/10.1088/2058-9565/ad37d4},
   DOI={10.1088/2058-9565/ad37d4},
   number={3},
   journal={Quantum Sci. Technol.},
   publisher={IOP Publishing},
   author={Cao, Shuxiang and Bakr, Mustafa and Campanaro, Giulio and Fasciati, Simone D and Wills, James and Lall, Deep and Shteynas, Boris and Chidambaram, Vivek and Rungger, Ivan and Leek, Peter},
   year={2024},
   month=apr, pages={035003} }

@article{qudit_encode1,
title = {Single qudit realization of the Deutsch algorithm using superconducting many-level quantum circuits},
journal = {Physics Letters A},
volume = {379},
number = {22},
pages = {1409-1413},
year = {2015},
issn = {0375-9601},
doi = {https://doi.org/10.1016/j.physleta.2015.03.023},
url = {https://www.sciencedirect.com/science/article/pii/S0375960115002753},
author = {E.O. Kiktenko and A.K. Fedorov and A.A. Strakhov and V.I. Man'ko}
}

@article{qudit_encode2,
  title = {Multilevel superconducting circuits as two-qubit systems: Operations, state preparation, and entropic inequalities},
  author = {Kiktenko, E. O. and Fedorov, A. K. and Man'ko, O. V. and Man'ko, V. I.},
  journal = {Phys. Rev. A},
  volume = {91},
  issue = {4},
  pages = {042312},
  numpages = {7},
  year = {2015},
  month = {Apr},
  publisher = {American Physical Society},
  doi = {10.1103/PhysRevA.91.042312},
  url = {https://link.aps.org/doi/10.1103/PhysRevA.91.042312}
}

@article{qudit_compile1,
  title = {Efficient Toffoli gates using qudits},
  author = {Ralph, T. C. and Resch, K. J. and Gilchrist, A.},
  journal = {Phys. Rev. A},
  volume = {75},
  issue = {2},
  pages = {022313},
  numpages = {5},
  year = {2007},
  month = {Feb},
  publisher = {American Physical Society},
  doi = {10.1103/PhysRevA.75.022313},
  url = {https://link.aps.org/doi/10.1103/PhysRevA.75.022313}
}

@article{qudit_compile2,
   author = {Fedorov, A. and Steffen, L. and Baur, M. and da Silva, M. P. and Wallraff, A.},
   title = {Implementation of a Toffoli gate with superconducting circuits},
   journal = {Nature},
   volume = {481},
   number = {7380},
   pages = {170-172},
   ISSN = {1476-4687},
   DOI = {10.1038/nature10713},
   url = {https://doi.org/10.1038/nature10713},
   year = {2012},
   type = {Journal Article}
}

@article{qudit_compile3,
  title = {Scalable quantum computing with qudits on a graph},
  author = {Kiktenko, E. O. and Nikolaeva, A. S. and Xu, Peng and Shlyapnikov, G. V. and Fedorov, A. K.},
  journal = {Phys. Rev. A},
  volume = {101},
  issue = {2},
  pages = {022304},
  numpages = {7},
  year = {2020},
  month = {Feb},
  publisher = {American Physical Society},
  doi = {10.1103/PhysRevA.101.022304},
  url = {https://link.aps.org/doi/10.1103/PhysRevA.101.022304}
}

@article{qudit_compile4,
  title = {Quantum Information Scrambling on a Superconducting Qutrit Processor},
  author = {Blok, M. S. and Ramasesh, V. V. and Schuster, T. and O'Brien, K. and Kreikebaum, J. M. and Dahlen, D. and Morvan, A. and Yoshida, B. and Yao, N. Y. and Siddiqi, I.},
  journal = {Phys. Rev. X},
  volume = {11},
  issue = {2},
  pages = {021010},
  numpages = {21},
  year = {2021},
  month = {Apr},
  publisher = {American Physical Society},
  doi = {10.1103/PhysRevX.11.021010},
  url = {https://link.aps.org/doi/10.1103/PhysRevX.11.021010}
}

@article{ qudit_photonics1 ,
   author = {Chi, Yulin and Huang, Jieshan and Zhang, Zhanchuan and Mao, Jun and Zhou, Zinan and Chen, Xiaojiong and Zhai, Chonghao and Bao, Jueming and Dai, Tianxiang and Yuan, Huihong and Zhang, Ming and Dai, Daoxin and Tang, Bo and Yang, Yan and Li, Zhihua and others},
   title = {A programmable qudit-based quantum processor},
   journal = {Nat. Commun.},
   volume = {13},
   number = {1},
   pages = {1166},
   ISSN = {2041-1723},
   DOI = {10.1038/s41467-022-28767-x},
   url = {https://doi.org/10.1038/s41467-022-28767-x},
   year = {2022},
   type = {Journal Article}
}

@article{ qudit_photonics2 ,
   author = {Bao, Jueming and Fu, Zhaorong and Pramanik, Tanumoy and Mao, Jun and Chi, Yulin and Cao, Yingkang and Zhai, Chonghao and Mao, Yifei and Dai, Tianxiang and Chen, Xiaojiong and Jia, Xinyu and Zhao, Leshi and Zheng, Yun and Tang, Bo and Li, Zhihua and others},
   title = {Very-large-scale integrated quantum graph photonics},
   journal = {Nat. Photonics},
   volume = {17},
   number = {7},
   pages = {573-581},
   ISSN = {1749-4893},
   DOI = {10.1038/s41566-023-01187-z},
   url = {https://doi.org/10.1038/s41566-023-01187-z},
   year = {2023},
   type = {Journal Article}
}

@article{ qudit_ions,
   author = {Ringbauer, Martin and Meth, Michael and Postler, Lukas and Stricker, Roman and Blatt, Rainer and Schindler, Philipp and Monz, Thomas},
   title = {A universal qudit quantum processor with trapped ions},
   journal = {Nat. Phys.},
   volume = {18},
   number = {9},
   pages = {1053-1057},
   ISSN = {1745-2481},
   DOI = {10.1038/s41567-022-01658-0},
   url = {https://doi.org/10.1038/s41567-022-01658-0},
   year = {2022},
   type = {Journal Article}
}

@article{ qudit_Rydberg ,
  title = {Hardware Efficient Quantum Simulation of Non-Abelian Gauge Theories with Qudits on Rydberg Platforms},
  author = {Gonz\'alez-Cuadra, Daniel and Zache, Torsten V. and Carrasco, Jose and Kraus, Barbara and Zoller, Peter},
  journal = {Phys. Rev. Lett.},
  volume = {129},
  issue = {16},
  pages = {160501},
  numpages = {8},
  year = {2022},
  month = {Oct},
  publisher = {American Physical Society},
  doi = {10.1103/PhysRevLett.129.160501},
  url = {https://link.aps.org/doi/10.1103/PhysRevLett.129.160501}
}

@article{ qudit_errormitigation,
   author = {Goss, Noah and Ferracin, Samuele and Hashim, Akel and Carignan-Dugas, Arnaud and Kreikebaum, John Mark and Naik, Ravi K. and Santiago, David I. and Siddiqi, Irfan},
   title = {Extending the computational reach of a superconducting qutrit processor},
   journal = {npj Quantum Inf.},
   volume = {10},
   number = {1},
   pages = {101},
   ISSN = {2056-6387},
   DOI = {10.1038/s41534-024-00892-z},
   url = {https://doi.org/10.1038/s41534-024-00892-z},
   year = {2024},
   type = {Journal Article}
}

@article{ qudit_errorcorrect1,
  title = {Enhanced Fault-Tolerant Quantum Computing in $d$-Level Systems},
  author = {Campbell, Earl T.},
  journal = {Phys. Rev. Lett.},
  volume = {113},
  issue = {23},
  pages = {230501},
  numpages = {5},
  year = {2014},
  month = {Dec},
  publisher = {American Physical Society},
  doi = {10.1103/PhysRevLett.113.230501},
  url = {https://link.aps.org/doi/10.1103/PhysRevLett.113.230501}
}

@article{qudit_errorcorrect2,
   author = {Brock, Benjamin L. and Singh, Shraddha and Eickbusch, Alec and Sivak, Volodymyr V. and Ding, Andy Z. and Frunzio, Luigi and Girvin, Steven M. and Devoret, Michel H.},
   title = {Quantum error correction of qudits beyond break-even},
   journal = {Nature},
   volume = {641},
   number = {8063},
   pages = {612-618},
   ISSN = {1476-4687},
   DOI = {10.1038/s41586-025-08899-y},
   url = {https://doi.org/10.1038/s41586-025-08899-y},
   year = {2025},
   type = {Journal Article}
}

@article{topo_xue,
  title = {Detecting Non-Bloch Topological Invariants in Quantum Dynamics},
  author = {Wang, Kunkun and Li, Tianyu and Xiao, Lei and Han, Yiwen and Yi, Wei and Xue, Peng},
  journal = {Phys. Rev. Lett.},
  volume = {127},
  issue = {27},
  pages = {270602},
  numpages = {7},
  year = {2021},
  month = {Dec},
  publisher = {American Physical Society},
  doi = {10.1103/PhysRevLett.127.270602},
  url = {https://link.aps.org/doi/10.1103/PhysRevLett.127.270602}
}

@article{q_sim_2,
  title = {Detecting Topological Invariants in Nonunitary Discrete-Time Quantum Walks},
  author = {Zhan, Xiang and Xiao, Lei and Bian, Zhihao and Wang, Kunkun and Qiu, Xingze and Sanders, Barry C. and Yi, Wei and Xue, Peng},
  journal = {Phys. Rev. Lett.},
  volume = {119},
  issue = {13},
  pages = {130501},
  numpages = {6},
  year = {2017},
  month = {Sep},
  publisher = {American Physical Society},
  doi = {10.1103/PhysRevLett.119.130501},
  url = {https://link.aps.org/doi/10.1103/PhysRevLett.119.130501}
}

@article{BS_2019,
doi = {10.1088/1367-2630/ab0610},
url = {https://doi.org/10.1088/1367-2630/ab0610},
year = {2019},
month = {may},
publisher = {IOP Publishing},
volume = {21},
number = {5},
pages = {055003},
author = {Muraleedharan, Gopikrishnan and Miyake, Akimasa and Deutsch, Ivan H},
title = {Quantum computational supremacy in the sampling of bosonic random walkers on a one-dimensional lattice},
journal = {New Journal of Physics}
}

@article{cold_atoms,
  title = {Topological Quantum Walks in Momentum Space with a Bose-Einstein Condensate},
  author = {Xie, Dizhou and Deng, Tian-Shu and Xiao, Teng and Gou, Wei and Chen, Tao and Yi, Wei and Yan, Bo},
  journal = {Phys. Rev. Lett.},
  volume = {124},
  issue = {5},
  pages = {050502},
  numpages = {6},
  year = {2020},
  month = {Feb},
  publisher = {American Physical Society},
  doi = {10.1103/PhysRevLett.124.050502},
  url = {https://link.aps.org/doi/10.1103/PhysRevLett.124.050502}
}

@article{qw_in_sc_1_phase_space,
  title = {Observing Topological Invariants Using Quantum Walks in Superconducting Circuits},
  author = {Flurin, E. and Ramasesh, V. V. and Hacohen-Gourgy, S. and Martin, L. S. and Yao, N. Y. and Siddiqi, I.},
  journal = {Phys. Rev. X},
  volume = {7},
  issue = {3},
  pages = {031023},
  numpages = {6},
  year = {2017},
  month = {Aug},
  publisher = {American Physical Society},
  doi = {10.1103/PhysRevX.7.031023},
  url = {https://link.aps.org/doi/10.1103/PhysRevX.7.031023}
}

@article{qw_in_sc_2_phase_space,
  title = {Direct Probe of Topological Invariants Using Bloch Oscillating Quantum Walks},
  author = {Ramasesh, V. V. and Flurin, E. and Rudner, M. and Siddiqi, I. and Yao, N. Y.},
  journal = {Phys. Rev. Lett.},
  volume = {118},
  issue = {13},
  pages = {130501},
  numpages = {6},
  year = {2017},
  month = {Mar},
  publisher = {American Physical Society},
  doi = {10.1103/PhysRevLett.118.130501},
  url = {https://link.aps.org/doi/10.1103/PhysRevLett.118.130501}
}

@article{qw_in_sc_2_PJW1D,
author = {Zhiguang Yan  and Yu-Ran Zhang  and Ming Gong  and Yulin Wu  and Yarui Zheng  and Shaowei Li  and Can Wang  and Futian Liang  and Jin Lin  and Yu Xu  and Cheng Guo  and Lihua Sun  and Cheng-Zhi Peng  and Keyu Xia  and Hui Deng  and others},
title = {Strongly correlated quantum walks with a 12-qubit superconducting processor},
journal = {Science},
volume = {364},
number = {6442},
pages = {753-756},
year = {2019},
doi = {10.1126/science.aaw1611},
URL = {https://www.science.org/doi/abs/10.1126/science.aaw1611},
eprint = {https://www.science.org/doi/pdf/10.1126/science.aaw1611}}

@article{qw_in_sc_3_PJW2D,
author = {Ming Gong  and Shiyu Wang  and Chen Zha  and Ming-Cheng Chen  and He-Liang Huang  and Yulin Wu  and Qingling Zhu  and Youwei Zhao  and Shaowei Li  and Shaojun Guo  and Haoran Qian  and Yangsen Ye  and Fusheng Chen  and Chong Ying  and Jiale Yu  and others},
title = {Quantum walks on a programmable two-dimensional 62-qubit superconducting processor},
journal = {Science},
volume = {372},
number = {6545},
pages = {948-952},
year = {2021},
doi = {10.1126/science.abg7812},
URL = {https://www.science.org/doi/abs/10.1126/science.abg7812},
eprint = {https://www.science.org/doi/pdf/10.1126/science.abg7812}}

@article{DTQW_2qbit,
  title = {Simulating Anderson localization via a quantum walk on a one-dimensional lattice of superconducting qubits},
  author = {Ghosh, Joydip},
  journal = {Phys. Rev. A},
  volume = {89},
  issue = {2},
  pages = {022309},
  numpages = {12},
  year = {2014},
  month = {Feb},
  publisher = {American Physical Society},
  doi = {10.1103/PhysRevA.89.022309},
  url = {https://link.aps.org/doi/10.1103/PhysRevA.89.022309}
}

@article{QW_search_1,
  title = {Spatial Search by Quantum Walk is Optimal for Almost all Graphs},
  author = {Chakraborty, Shantanav and Novo, Leonardo and Ambainis, Andris and Omar, Yasser},
  journal = {Phys. Rev. Lett.},
  volume = {116},
  issue = {10},
  pages = {100501},
  numpages = {5},
  year = {2016},
  month = {Mar},
  publisher = {American Physical Society},
  doi = {10.1103/PhysRevLett.116.100501},
  url = {https://link.aps.org/doi/10.1103/PhysRevLett.116.100501}
}

@article{QW_search_2,
  title = {Quadratic Speedup for Spatial Search by Continuous-Time Quantum Walk},
  author = {Apers, Simon and Chakraborty, Shantanav and Novo, Leonardo and Roland, J\'er\'emie},
  journal = {Phys. Rev. Lett.},
  volume = {129},
  issue = {16},
  pages = {160502},
  numpages = {6},
  year = {2022},
  month = {Oct},
  publisher = {American Physical Society},
  doi = {10.1103/PhysRevLett.129.160502},
  url = {https://link.aps.org/doi/10.1103/PhysRevLett.129.160502}
}

@article{Graph_2,
author = {Xiaogang Qiang  and Yizhi Wang  and Shichuan Xue  and Renyou Ge  and Lifeng Chen  and Yingwen Liu  and Anqi Huang  and Xiang Fu  and Ping Xu  and Teng Yi  and Fufang Xu  and Mingtang Deng  and Jingbo B. Wang  and Jasmin D. A. Meinecke  and others },
title = {Implementing graph-theoretic quantum algorithms on a silicon photonic quantum walk processor},
journal = {Sci. Adv.},
volume = {7},
number = {9},
pages = {eabb8375},
year = {2021},
doi = {10.1126/sciadv.abb8375},
URL = {https://www.science.org/doi/abs/10.1126/sciadv.abb8375},
}

@article{DTQW_circuit,
  title = {Efficient quantum circuit implementation of quantum walks},
  author = {Douglas, B. L. and Wang, J. B.},
  journal = {Phys. Rev. A},
  volume = {79},
  issue = {5},
  pages = {052335},
  numpages = {5},
  year = {2009},
  month = {May},
  publisher = {American Physical Society},
  doi = {10.1103/PhysRevA.79.052335},
  url = {https://link.aps.org/doi/10.1103/PhysRevA.79.052335}
}

@article{pra2010,
  title = {Exploring topological phases with quantum walks},
  author = {Kitagawa, Takuya and Rudner, Mark S. and Berg, Erez and Demler, Eugene},
  journal = {Phys. Rev. A},
  volume = {82},
  issue = {3},
  pages = {033429},
  numpages = {9},
  year = {2010},
  month = {Sep},
  publisher = {American Physical Society},
  doi = {10.1103/PhysRevA.82.033429},
  url = {https://link.aps.org/doi/10.1103/PhysRevA.82.033429}
}

@article{prb2012,
  title = {Symmetries, topological phases, and bound states in the one-dimensional quantum walk},
  author = {Asb\'oth, J. K.},
  journal = {Phys. Rev. B},
  volume = {86},
  issue = {19},
  pages = {195414},
  numpages = {9},
  year = {2012},
  month = {Nov},
  publisher = {American Physical Society},
  doi = {10.1103/PhysRevB.86.195414},
  url = {https://link.aps.org/doi/10.1103/PhysRevB.86.195414}
}

@article{pra2023,
  title = {Topological invariants in quantum walks},
  author = {Grudka, Andrzej and Karczewski, Marcin and Kurzy\ifmmode \acute{n}\else \'{n}\fi{}ski, Pawe\l{} and W\'ojcik, Jan and W\'ojcik, Antoni},
  journal = {Phys. Rev. A},
  volume = {107},
  issue = {3},
  pages = {032201},
  numpages = {7},
  year = {2023},
  month = {Mar},
  publisher = {American Physical Society},
  doi = {10.1103/PhysRevA.107.032201},
  url = {https://link.aps.org/doi/10.1103/PhysRevA.107.032201}
}

@article{splitstep,
  title = {Eigenvalue measurement of topologically protected edge states in split-step quantum walks},
  author = {Nitsche, Thomas and Geib, Tobias and Stahl, Christoph and Lorz, Lennart and Cedzich, Christopher and Barkhofen, Sonja and Werner, Reinhard F and Silberhorn, Christine},
  journal = {New J. Phys.},
  volume = {21},
  number = {4},
  pages = {043031},
  year = {2019},
  month = {apr},
  publisher = {IOP Publishing},
  doi = {10.1088/1367-2630/ab12fa},
  url = {https://dx.doi.org/10.1088/1367-2630/ab12fa}
}

@article{UD_1,
  title = {Unidirectional quantum walks: Evolution and exit times},
  author = {Montero, Miquel},
  journal = {Phys. Rev. A},
  volume = {88},
  issue = {1},
  pages = {012333},
  numpages = {9},
  year = {2013},
  month = {Jul},
  publisher = {American Physical Society},
  doi = {10.1103/PhysRevA.88.012333},
  url = {https://link.aps.org/doi/10.1103/PhysRevA.88.012333}
}

@article{UD_2,
  title = {Quantum state engineering using one-dimensional discrete-time quantum walks},
  author = {Innocenti, Luca and Majury, Helena and Giordani, Taira and Spagnolo, Nicol\`o and Sciarrino, Fabio and Paternostro, Mauro and Ferraro, Alessandro},
  journal = {Phys. Rev. A},
  volume = {96},
  issue = {6},
  pages = {062326},
  numpages = {11},
  year = {2017},
  month = {Dec},
  publisher = {American Physical Society},
  doi = {10.1103/PhysRevA.96.062326},
  url = {https://link.aps.org/doi/10.1103/PhysRevA.96.062326}
}

@article{2Dwalk,
  title = {Mimicking the Probability Distribution of a Two-Dimensional Grover Walk with a Single-Qubit Coin},
  author = {Di Franco, C. and Mc Gettrick, M. and Busch, Th.},
  journal = {Phys. Rev. Lett.},
  volume = {106},
  issue = {8},
  pages = {080502},
  numpages = {4},
  year = {2011},
  month = {Feb},
  publisher = {American Physical Society},
  doi = {10.1103/PhysRevLett.106.080502},
  url = {https://link.aps.org/doi/10.1103/PhysRevLett.106.080502}
}

@article{topo2D_xue,
  title = {Observation of Topologically Protected Edge States in a Photonic Two-Dimensional Quantum Walk},
  author = {Chen, Chao and Ding, Xing and Qin, Jian and He, Yu and Luo, Yi-Han and Chen, Ming-Cheng and Liu, Chang and Wang, Xi-Lin and Zhang, Wei-Jun and Li, Hao and You, Li-Xing and Wang, Zhen and Wang, Da-Wei and Sanders, Barry C. and Lu, Chao-Yang and others},
  journal = {Phys. Rev. Lett.},
  volume = {121},
  issue = {10},
  pages = {100502},
  numpages = {6},
  year = {2018},
  month = {Sep},
  publisher = {American Physical Society},
  doi = {10.1103/PhysRevLett.121.100502},
  url = {https://link.aps.org/doi/10.1103/PhysRevLett.121.100502}
}

@article{a4,
   author = {Xiao, L. and Zhan, X. and Bian, Z. H. and Wang, K. K. and Zhang, X. and Wang, X. P. and Li, J. and Mochizuki, K. and Kim, D. and Kawakami, N. and Yi, W. and Obuse, H. and Sanders, B. C. and Xue, P.},
   title = {Observation of topological edge states in parity–time-symmetric quantum walks},
   journal = {Nat. Phys.},
   volume = {13},
   number = {11},
   pages = {1117-1123},
   ISSN = {1745-2481},
   DOI = {10.1038/nphys4204},
   url = {https://doi.org/10.1038/nphys4204},
   year = {2017},
   type = {Journal Article}
}

@article{q_alg_1,
  title = {Quantum random-walk search algorithm},
  author = {Shenvi, Neil and Kempe, Julia and Whaley, K. Birgitta},
  journal = {Phys. Rev. A},
  volume = {67},
  issue = {5},
  pages = {052307},
  numpages = {11},
  year = {2003},
  month = {May},
  publisher = {American Physical Society},
  doi = {10.1103/PhysRevA.67.052307},
  url = {https://link.aps.org/doi/10.1103/PhysRevA.67.052307}
}

@article{q_alg_3,
   author = {Jeong, Youn-Chang and Di Franco, Carlo and Lim, Hyang-Tag and Kim, M. S. and Kim, Yoon-Ho},
   title = {Experimental realization of a delayed-choice quantum walk},
   journal = {Nat. Commun.},
   volume = {4},
   number = {1},
   pages = {2471},
   ISSN = {2041-1723},
   DOI = {10.1038/ncomms3471},
   url = {https://doi.org/10.1038/ncomms3471},
   year = {2013},
   type = {Journal Article}
}

@article{q_alg_4,
author = {Zi-Yu Shi and Hao Tang and Zhen Feng and Yao Wang and Zhan-Ming Li and Jun Gao and Yi-Jun Chang and Tian-Yu Wang and Jian-Peng Dou and Zhe-Yong Zhang and Zhi-Qiang Jiao and Wen-Hao Zhou and Xian-Min Jin},
journal = {Optica},
number = {6},
pages = {613--618},
publisher = {Optica Publishing Group},
title = {Quantum fast hitting on glued trees mapped on a photonic chip},
volume = {7},
month = {Jun},
year = {2020},
doi = {10.1364/OPTICA.388451},
}

@article{q_sim_1,
   author = {Crespi, Andrea and Osellame, Roberto and Ramponi, Roberta and Giovannetti, Vittorio and Fazio, Rosario and Sansoni, Linda and De Nicola, Francesco and Sciarrino, Fabio and Mataloni, Paolo},
   title = {Anderson localization of entangled photons in an integrated quantum walk},
   journal = {Nat. Photonics},
   volume = {7},
   number = {4},
   pages = {322-328},
   ISSN = {1749-4893},
   DOI = {10.1038/nphoton.2013.26},
   url = {https://doi.org/10.1038/nphoton.2013.26},
   year = {2013},
   type = {Journal Article}
}

@article{q_sim_3,
   author = {Peruzzo, A. and Lobino, M. and Matthews, J. C. and Matsuda, N. and Politi, A. and Poulios, K. and Zhou, X. Q. and Lahini, Y. and Ismail, N. and Wörhoff, K. and Bromberg, Y. and Silberberg, Y. and Thompson, M. G. and JL, O. Brien},
   title = {Quantum walks of correlated photons},
   journal = {Science},
   volume = {329},
   number = {5998},
   pages = {1500-3},
   ISSN = {0036-8075},
   DOI = {10.1126/science.1193515},
   year = {2010},
   type = {Journal Article}
}

@article{u_c_1,
  title = {Universal Computation by Quantum Walk},
  author = {Childs, Andrew M.},
  journal = {Phys. Rev. Lett.},
  volume = {102},
  issue = {18},
  pages = {180501},
  numpages = {4},
  year = {2009},
  month = {May},
  publisher = {American Physical Society},
  doi = {10.1103/PhysRevLett.102.180501},
  url = {https://link.aps.org/doi/10.1103/PhysRevLett.102.180501}
}

@article{u_c_2,
  title = {Universal quantum computation by discontinuous quantum walk},
  author = {Underwood, Michael S. and Feder, David L.},
  journal = {Phys. Rev. A},
  volume = {82},
  issue = {4},
  pages = {042304},
  numpages = {6},
  year = {2010},
  month = {Oct},
  publisher = {American Physical Society},
  doi = {10.1103/PhysRevA.82.042304},
  url = {https://link.aps.org/doi/10.1103/PhysRevA.82.042304}
}

@article{u_c_3,
  author = {Andrew M. Childs  and David Gosset  and Zak Webb },
  title = {Universal Computation by Multiparticle Quantum Walk},
  journal = {Science},
  volume = {339},
  number = {6121},
  pages = {791-794},
  year = {2013},
  doi = {10.1126/science.1229957},
  URL = {https://www.science.org/doi/abs/10.1126/science.1229957},
  eprint = {https://www.science.org/doi/pdf/10.1126/science.1229957}}

@article{cp_1,
  title = {Tunable Coupling Scheme for Implementing High-Fidelity Two-Qubit Gates},
  author = {Yan, Fei and Krantz, Philip and Sung, Youngkyu and Kjaergaard, Morten and Campbell, Daniel L. and Orlando, Terry P. and Gustavsson, Simon and Oliver, William D.},
  journal = {Phys. Rev. Applied},
  volume = {10},
  issue = {5},
  pages = {054062},
  numpages = {9},
  year = {2018},
  month = {Nov},
  publisher = {American Physical Society},
  doi = {10.1103/PhysRevApplied.10.054062},
  url = {https://link.aps.org/doi/10.1103/PhysRevApplied.10.054062}
}

@article{cp_2,
  title = {Tunable Coupling Architecture for Fixed-Frequency Transmon Superconducting Qubits},
  author = {Stehlik, J. and Zajac, D. M. and Underwood, D. L. and Phung, T. and Blair, J. and Carnevale, S. and Klaus, D. and Keefe, G. A. and Carniol, A. and Kumph, M. and Steffen, Matthias and Dial, O. E.},
  journal = {Phys. Rev. Lett.},
  volume = {127},
  issue = {8},
  pages = {080505},
  numpages = {6},
  year = {2021},
  month = {Aug},
  publisher = {American Physical Society},
  doi = {10.1103/PhysRevLett.127.080505},
  url = {https://link.aps.org/doi/10.1103/PhysRevLett.127.080505}
}

@article{cp_3,
  title = {Floating Tunable Coupler for Scalable Quantum Computing Architectures},
  author = {Sete, Eyob A. and Chen, Angela Q. and Manenti, Riccardo and Kulshreshtha, Shobhan and Poletto, Stefano},
  journal = {Phys. Rev. Applied},
  volume = {15},
  issue = {6},
  pages = {064063},
  numpages = {12},
  year = {2021},
  month = {Jun},
  publisher = {American Physical Society},
  doi = {10.1103/PhysRevApplied.15.064063},
  url = {https://link.aps.org/doi/10.1103/PhysRevApplied.15.064063}
}

@article{cp_4,
  title = {Long-Distance Transmon Coupler with cz-Gate Fidelity above $99.8\mathrm{\%}$},
  author = {Marxer, Fabian and Veps\"al\"ainen, Antti and Jolin, Shan W. and Tuorila, Jani and Landra, Alessandro and Ockeloen-Korppi, Caspar and Liu, Wei and Ahonen, Olli and Auer, Adrian and Belzane, Lucien and Bergholm, Ville and Chan, Chun Fai and Chan, Kok Wai and Hiltunen, Tuukka and Hotari, Juho and others},
  journal = {PRX Quantum},
  volume = {4},
  issue = {1},
  pages = {010314},
  numpages = {23},
  year = {2023},
  month = {Feb},
  publisher = {American Physical Society},
  doi = {10.1103/PRXQuantum.4.010314},
  url = {https://link.aps.org/doi/10.1103/PRXQuantum.4.010314}
}

@article{transmon,
  title = {Charge-insensitive qubit design derived from the Cooper pair box},
  author = {Koch, Jens and Yu, Terri M. and Gambetta, Jay and Houck, A. A. and Schuster, D. I. and Majer, J. and Blais, Alexandre and Devoret, M. H. and Girvin, S. M. and Schoelkopf, R. J.},
  journal = {Phys. Rev. A},
  volume = {76},
  issue = {4},
  pages = {042319},
  numpages = {19},
  year = {2007},
  month = {Oct},
  publisher = {American Physical Society},
  doi = {10.1103/PhysRevA.76.042319},
  url = {https://link.aps.org/doi/10.1103/PhysRevA.76.042319}
}

@misc{SupplementalMaterial,
  author = {},
  title = {},
  howpublished = {See Supplemental Material at \url{https://doi.org/[THIS-DOI]} for the supporting information of additional data, experimental and simulation methods, and extended discussions, which includes Refs. 56-60.},
  note = {},
  year = {}
}

@article{QW_EXPSUP,
   author = {Balasubramanian, Shankar and Li, Tongyang and Harrow, Aram W.},
   title = {Exponential Speedups for Quantum Walks in Random Hierarchical Graphs},
   journal = {Communications in Mathematical Physics},
   volume = {406},
   number = {9},
   pages = {209},
   ISSN = {1432-0916},
   DOI = {10.1007/s00220-025-05370-x},
   url = {https://doi.org/10.1007/s00220-025-05370-x},
   year = {2025},
   type = {Journal Article}
}

\end{document}